\documentclass[a4paper]{article}

\usepackage{INTERSPEECH2020}
\usepackage[labelfont=bf]{caption}
\usepackage{amsmath}
\usepackage{booktabs}
\usepackage{multirow}
\usepackage{float}

\title{MASKED PRE-TRAINED ENCODER BASE ON JOINT CTC-TRANSFORMER}
\name{Lu Liu, Yiheng Huang}
\address{
 Tsinghua-Berkeley Shenzhen Institute, Tsinghua University\\
  Tencent AI lab}
\email{liulu18@mails.tsinghua.edu.cn, arnoldhuang@tencent.com}

\begin{document}

\maketitle
\begin{abstract}
This study\footnote{The work was accomplished during the internship in Tencent AI lab} addresses semi-supervised 
acoustic modeling, i.e. attaining high-level representations from unsupervised audio data and fine-tuning the parameters of pre-trained model with supervised data. The proposed approach adopts a two-stage training framework, consisting of masked pretained encoder (MPE) and Joint CTC-Transformer (JCT). In the MPE framework, part of input frames are masked and reconstructed after the encoder with massive unsupervised data. In JCT framework, compared with original Transformer, acoustic features are applied as input instead of plain text. CTC loss performs as the prediction target on top of the encoder, 
and decoder blocks remain unchanged. This paper presents a comparison between two-stage training method and the fully supervised JCT. In addition, this paper investigates the our approach's robustness against different volumns of training data. Experiments on the two-stage training method deliver much better performance than fully supervised model. The word error rate (WER) with two-stage training which only exploits 30\% of WSJ labeled data achieves 17\% reduction than which trained by 50\% of WSJ in a fully supervised way. Moreover, increasing unlabeled data for MPE from WSJ (81h) to Librispeech (960h) attains about 22\% WER reduction.
\end{abstract}
\noindent\textbf{Index Terms}: unsupervised learning, transformer, ASR

\section{Introduction}
Data labeling is quite time consuming and costly. However, most existing high-performance automatic speech recognition (ASR) systems require substantial speech data paired with transcriptions. In order to make use of unlabeled data, we propose a semi-supervised training method which combines unsupervised pre-training \cite{Barlow89} and supervised training together. Our pre-training method is mainly inspired by the unsupervised pre-training approach in natural language processing (NLP) tasks, especially the most representative work
\textit{BERT} \cite{bert}, which has refreshed the state-of-the-art of dozens of NLP tasks. Our supervised training structure JCT is inspired by CTC \cite{graves2006} and Transformer \cite{transformer}. CTC is a commonly used end-to-end speech recognition loss function in recurrent neural network (RNN) based models, such as \cite{deepspeech2}. Transformer is an encoder-decoder model that is widely applied in sequence to sequence tasks, such as \cite{dong2018speech,zhou2018syllable,zhou2018,zhou2018multilingual}. Transformer shows much better performance in parallel computing and long sequences modeling, compared to RNN based models \cite{Graves2012,Sak2017}. In JCT model, CTC simply acts as an auxiliary function in supervised training process. Consequently, we exploit a shared encoder to train CTC and transformer jointly during the supervised training process. 

\begin{figure}[htbp]
  \centering
  \includegraphics[width=0.9\linewidth]{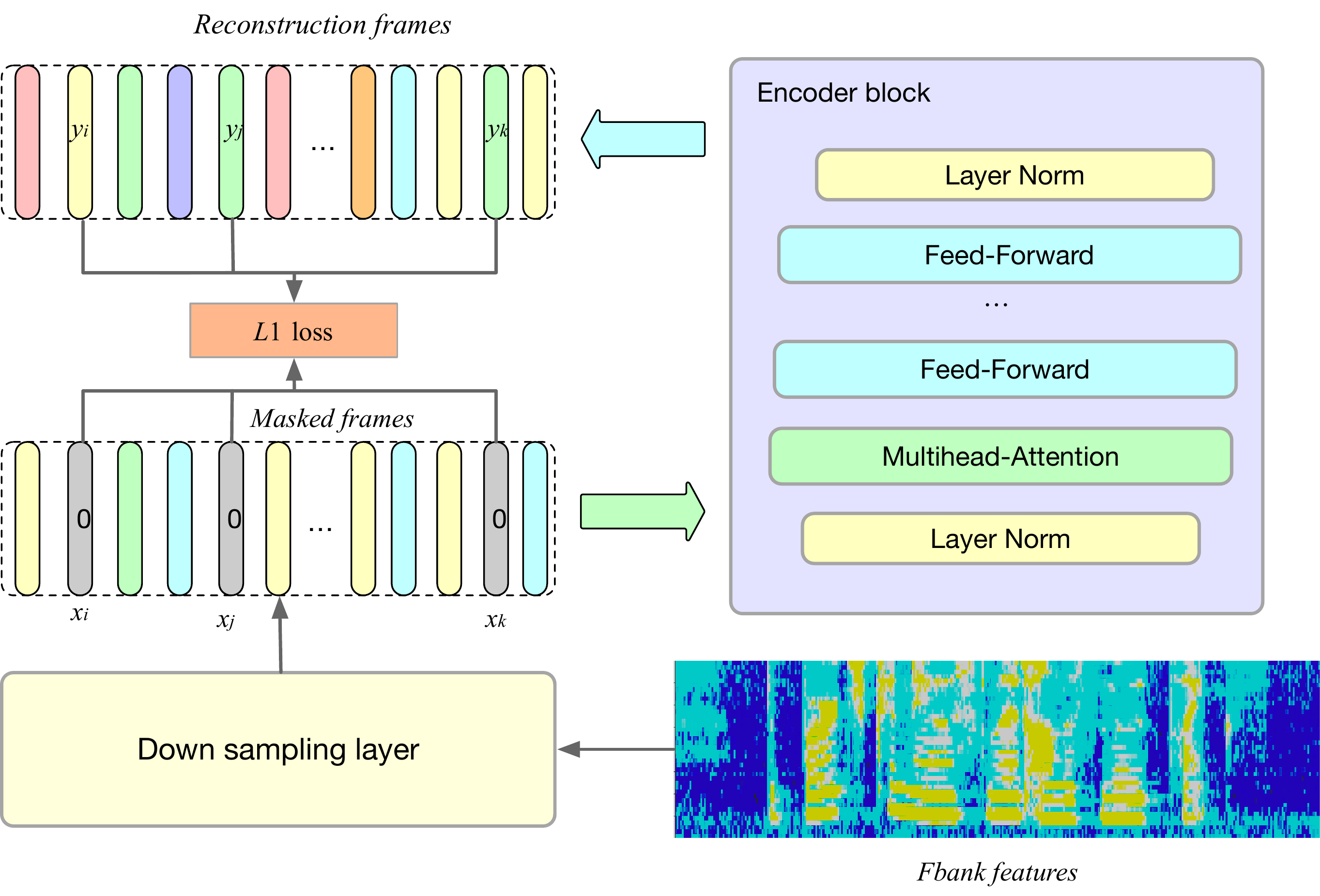}
  \caption{\rm The structure of masked pre-trained model}
  \label{fig:pre_train}
\end{figure}

\textit{BERT} is a pre-trained language model (LM) consisting of masked LM task and next sentence prediction task. These two tasks respectively capture word-level representation and sentence level representation. While for ASR tasks, audio examples in training dataset have no relationship and lack of contextual coherent information between each other. Thus we abandon next sentence prediction task in our pre-trained model. Meanwhile, in masked LM task, \textit{BERT} generates masks for original text data with special mask token ([MASK]). However acoustic features such as Mel-Frequency Cepstral Coefficients (MFCC) features and log-mel filter bank (FBANK) features \cite{MFCC} are much more complex than plain text features - the unclear alignment between acoustic frames and their transcriptions make it impossible to mask raw audio data in semantic level. Naturally, we mask the frames in neural networks. The implementation structure of the pre-trained model is a deep bidirectional transformer. Figure \ref{fig:pre_train} demonstrates the structure of masked pre-trained model. We use FBANK features as input and mask 15\% of the input down-sampled frames. Different from the conventional approach, our pre-training process exploits the information from both past and future frames to establish present masked frame, frames are then reconstructed as context representations. 
 As a kind of novel high-level representations of acoustic features, these representations are less-noisy that can be conveniently applied to dispose downstream speech processing tasks, such as speaker recognition, speaker verification and speech enhancement. In our paper, we explored the downstream task of low resource speech recognition to show that masked pre-trained encoder (MPE) is capable of improving supervised learning.
 
 Existing work related to unsupervised pre-training in speech recognition mainly focuses on the approaches of extracting high-level acoustic representations. Wave2vec \cite{wave2vec} proposed an unsupervised pre-training method by learning from the original audio signals rather than FBANK features, optimized by the noise contrastive estimation (NCE) of a binary classification task. Contrastive Predictive Coding (CPC) \cite{CPC} compresses the higher-dimensional data into a more compact potentially embedded space where conditional prediction is easier to be modeled. Then the researchers construct powerful autoregressive models in this potential space to make multi-step future predictions. CPC is also optimized by NCE. Compared with CPC, Autoregressive Predictive Coding (APC) \cite{APC} mainly focus on predicting the spectrum of a future frame rather than a wave sample, which appears like language model. The researchers use RNN based model to reconstruct temporal frame with information from its past frames, and the optimization target is reconstruction discrepancy.  

Recently published literature Deep Contextualized Acoustic Representations (DeCoAR) \cite{Shaoshi2019} introduces a new representation learning method. In this paper, a temporal slice of filterbank features from past and future context frames are reconstructed, the model is implemented by bi-directional LSTM networks and optimization target is reconstrction error. Mockingjay \cite{mockingjay} proposes a speech representation learning approach as \textit{BERT}, where bidirectional Transformer encoders are pre-trained on a large amount of unlabeled speech data and these representations are applied to a wide range of downstream tasks in ASR. Unlike their work, we mask the frames after down sampling layer while Mockingjay directly masks FBANK features before down sampling layer. We also expolit different down-samling method and distinctive supervised learning strategy from Mockingjay.
\section{Methodology}
The general framework of our proposed approach is illustrated in Figure 2. Given untranscribed speech data of a corpus, MPE is pretrained in the first stage. Next, JCT model is trained with a small amount of supervisd data. Pretrained features created by the MPE perform as high level representations.  Two different fine-tuning approaches are adopted consisting of directly fine-tuning and frozen fine-tuning.  The whole pipeline of our approach will be compared with a system consisting of only supervised data. Moreover, increasing the training data in MPE process will also be compared during experimental design. 
\subsection{Unsupervised pre-trained encoder}
To encode temporal information, current methods such as CPC and wave2vec adopt autoregressive models, which limits the potential of speech representation learning, and also decrease the training speed. By contrast, we leverage bi-directional transformer to reconstruct masked frame with its past and future frames.
\begin{figure}[htbp]
  \centering
  \includegraphics[width=\linewidth]{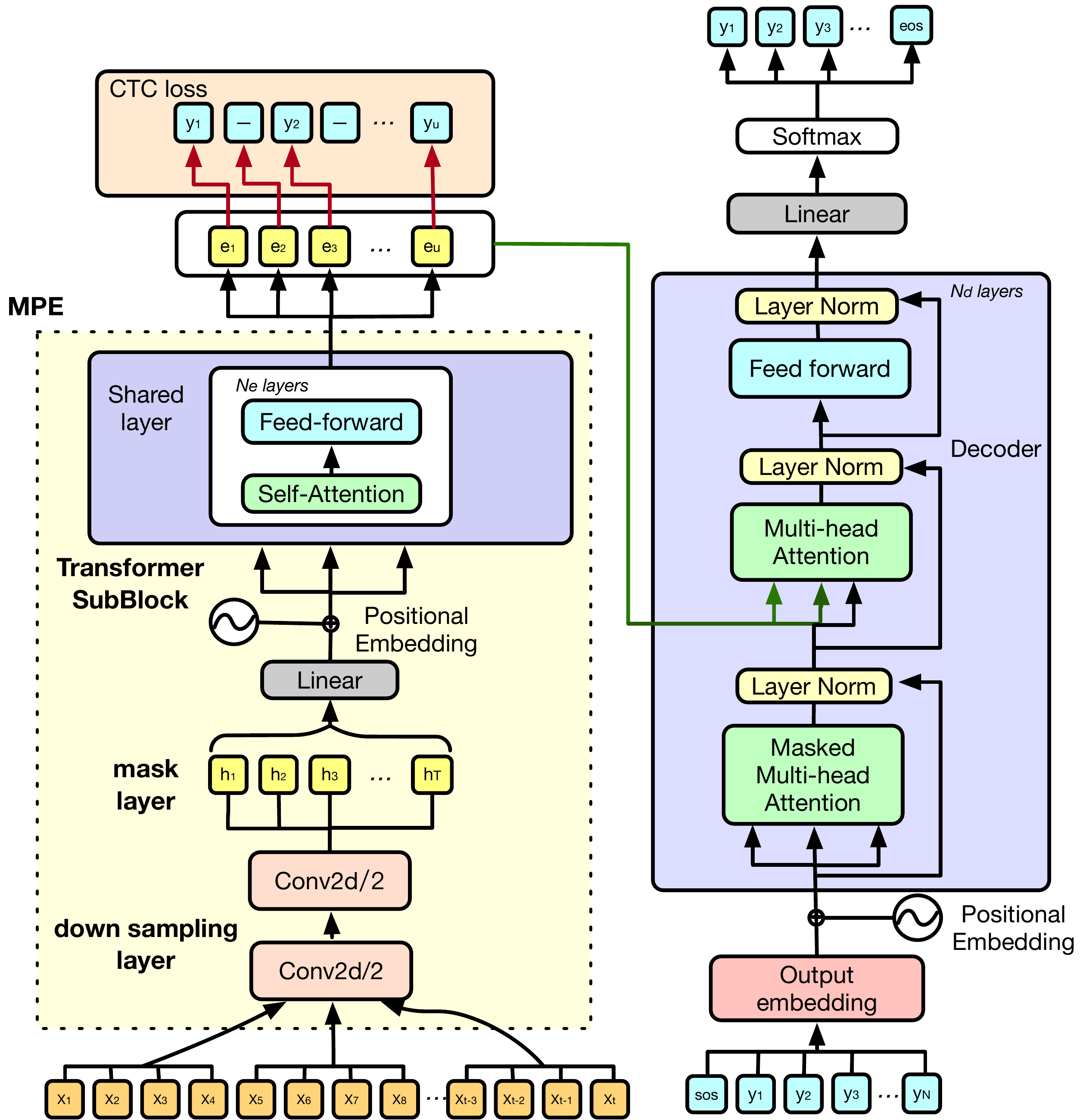}
  \caption{\rm{The structure of the semi-supervised JCT}}
  \label{fig:model structure}
\end{figure}
The structure of MPE is illustrated on the left of Figure \ref{fig:model structure}, which consists of three parts: down sampling layer, mask layer, bi-directional Transformer block. 
Considering faster calculation in training process, we place a downsampling layer before transformer block to exploit the structure locality of acoustic spectrograms \cite{mohamed2019}. The down-sampling layer consists of two convolutional neural networks (CNN). And the striding methods in both two CNN reduce the length of feature map to a quarter of its original length.
After that, we add a linear projection layer to reshape the dimensions of features to fit for the dimension of the transformer input. 
Then we present a random mask after the linear hidden layer with following setup: 15\% of the input frames need to be masked.  The chosen frames are replaced with zero vectors for 80\% of the time, with frames from random
positions 10\% of the time, and kept to be the same in the rest of the time. We also add sinusoidal positional embedding to the input features. The bi-directional Transformer block consists of $N_{e}$ layers of modules that can be stacked on top of each other multiple times. Each module composed of two sub-layers: multi-head attention layer and feed-forward layer, 
each sub-layer has a residual connection around it, and is followed by a layer-normalization step. 
Given  $t$ as length of input features, $T$ as length of MPE output sequences. \textbf x = $(x_1,x_2,...,x_t)$, \textbf e = $(e_1,e_2,...,e_T)$ respectively represent input features and reconstructed representations. \textbf h = $(h_1,h_2,...,h_T)$ is the masked down-sampled acoustic features.
\begin{align}
 	& \rm \textbf h = Mask(Conv(Conv(\textbf x))\\
 	&\rm \textbf e =  \textbf h + SubBlock(\textbf h)
\end{align}
Thus, the reconstruction discrepancy can be depicted as:
\begin{equation}
    {\mathcal{L}}_{pre}= \sum_{i=1}^{T}\left | h _{i} - e_{i} \right |
\end{equation}
The element in loss function merely contains the frames that keep unchanged rather than those that have been masked.
\subsection{Supervised encoder-decoder}
We exploit JCT in down stream supervised tasks. Based on the encoder-decoder structure of Transformer, on top of the encoder, CTC loss has been added as the prediction target. Pure CTC-based model always works together with a language model because of its independent assumption on the output elements. Pure attention-based model is hard to learn from scratch due to the sensitivity of attention mechanism \cite{kim2017joint}. Consequently, we integrate CTC with Transformer through the shared encoder MPE. In our experiments, attention mechanism tends to be impacted by noise,
which leads to bad alignment for modeling. The forward-backward algorithm of CTC loss precisely enforces monotonic alignment between input and output. Thus JCT performs more robust than the purely attention based model and CTC based model. 
Moreover, using CTC as an auxiliary optimization function speeds up the process of estimating the desired alignment than solely depending on data-driven attention methods.  

The right part of Figure \ref{fig:model structure} illustrates the structure of decoder, which is similar to the encoder, except for the masked multi-head attention module. To prevent adding future information and preserve the auto-regressive manner in the decoder, the masks in the masked multi-head attention module sweep out all values of illegal connections. This masking of the sequence can be achieved in parallel using an elementwise product with a triangular binary matrix. \textbf y = $(y_1,y_2,...,y_N)$ represent the transcriptions of audio data.
\begin{footnotesize}
\begin{align}
	&{\mathcal{L}}_{\rm CTC} = \sum_{(\rm{\textbf {x,y}})}-log(P (\rm{\textbf y|\textbf x})) \\
	{\mathcal{L}}_{\rm{Attention}}=-&\log P(\textbf {{y$\mid$ x}})=-\sum_{u}\log P(y_u^*|x,y_{[1:u-1]}^*)
\end{align}
\end{footnotesize}
where $y_{[1:u-1]}^*$ is the ground truth of the previous words. The joint training method of CTC with Transformer works as:
\begin{equation}
 {\mathcal{L}}_{\rm{JCT}}=\alpha {\mathcal{L}}_{\rm{CTC}} +(1-\alpha ){\mathcal{L}}_{\rm{Attention}}
\end{equation}
$\alpha$ is a hyper-parameter : $ 0\leqslant \alpha\leqslant 1$.
\subsection {Fine-tuning methods}
We leverage massive unsupervised audio data to train MPE. The training process won't stop until the result in validation dataset converges to the threshold of pre-training loss. After the completion of pre-training, we propose two approaches for the fine-tuning stage: 
\begin{itemize}
\item[*] Directly fine-tuning: Initialize the trainable parameters of encoder in JCT with the results we get from the pre-training process of MPE, then use labeled data to optimize the supervised joint loss function (JCT).
\item[*] Frozen fine-tuning: MPE provides more implicit and high-level representations than FBANK features. In the fine-tuning process, it performs better when we freeze the encoder, and only train the parameters of JCT decoder.
Specifically, frozen fine-tuning
 means remove the parameters of MPE from the trainable parameters of JCT. After the accomplishment of decoder training process, for better performance, we can train the whole structure in a supervised manner for a few epochs.
\end{itemize}
In our experiments, we have explored both fine-tuning methods, with the latter showing much better performance than the former. Essentially, the former fine-tuning method is a simple initialization of encoder in the supervised training stage, integrated with randomly initialized decoder will lose some information we attain from unlabeled data. If we choose the former method, the difference between directly fine-tuning method and totally supervised training method will be very small. While the latter one thoroughly uses the representation from massive unsupervised data, it shows much lower word error rate (WER) than totally supervised training in a low resource setting. The result are demonstrated in section 5.
\section{Experiments}
\subsection{Datasets}
We carried out experiments on LibriSpeech corpus and wall street journal (WSJ) corpus \cite{wsjcorpus}. For LibriSpeech \cite{librispeech} which contains 960 hours training audio data, we used the entire dataset to train MPE for high-level feature extraction. In the fine-tuning process, we exploited \textit{train-clean-100} and \textit{train-clean-360} for supervised training, \textit{dev-clean} for validation and \textit{test-clean} for evaluation. As for WSJ, the models were training on \textit{si284} which includes about 81 hours audio data, validating on \textit{dev93} and evaluating on \textit{eval92}. To evaluate the effect of MPE, we leverage the whole dataset for pre-training while one third, a half and the entire dataset are respectively used for supervised training. Meanwhile, an ideal feature extractor should extract representations that generalize to datasets of different domains. Thus, to examine the robustness of shifting in domains, we firstly trained MPE on LibriSpeech, then fine-tuned it to JCT with WSJ 81 hours supervised data. We choose totally supervised training on JCT as our baseline.
\subsection{Experiment setups}
The input acoustic features are 80-dimensional filterbanks extracted with a hop size of 10ms and a window size of 25ms, extended with temporal first and second order differences. Feature extraction also includes per-speaker mean subtraction and variance normalization \cite{dong2018speech}. The MPE consists of 2 CNN layers with RELU activation function and a stack of 12 encoder blocks, CNN has stride size 2 and kernel size 3 for down sampling. The channels of first layer is 64, next layer has twice as many channels as the previous one. For encoder blocks, each block contains two sub-layers: feed-forward layer (FFL) and self-attention layer (SAL), the dimension of FFL is 2048, as for SAL, the attention heads is 4 and dimension of embedding is 512. The SAL and FFL in the decoder obeys the same configuration, while the number of decoder stacked blocks is set to 6. 
We used Adam optimizer with default parameter configuration in both two-stage training. Especially in supervised training process, we applied warming up method to vary the learning rate in the whole training process with Noam learning strategy.
\begin{footnotesize}
\begin{equation}
{\rm lr} = k\ast  d^{-0.5} \ast {\rm min}(n^{-0.5},n\ast warmup^{-1.5} )
\end{equation}
\end{footnotesize}
\textit{k}, \textit{d}, \textit{n}, \textit{warmup} respectively refers to a tunable hyper-parameter, model dimension, training step, total warming up steps. The learning rate increased in start warming up \textit{n} steps and decreased after the peak of lr. In our experiments,  warming up steps \textit{n} = 25000, hyper-parameter \textit{k} = 10. To avoid over-fitting, label smoothing strategy which was proposed in \cite{label_smoothing} was also applied in the training process, and the label smoothing weight is set as 0.1. Meanwhile, both of residual dropout and attention dropout \cite{dropout} were set to 0.1. Moreover, we also used SortaGrad \cite{SortaGrad} method in the first training epoch for faster convergence and less noise inference. Apart from above configuration, for the multi-task training process, the hyper-pramater $\alpha$ is set as 0.3. 
\section{Results}
\subsection {Pre-training results}
\begin{figure}[htbp]
  \centering
  \includegraphics[width=\linewidth]{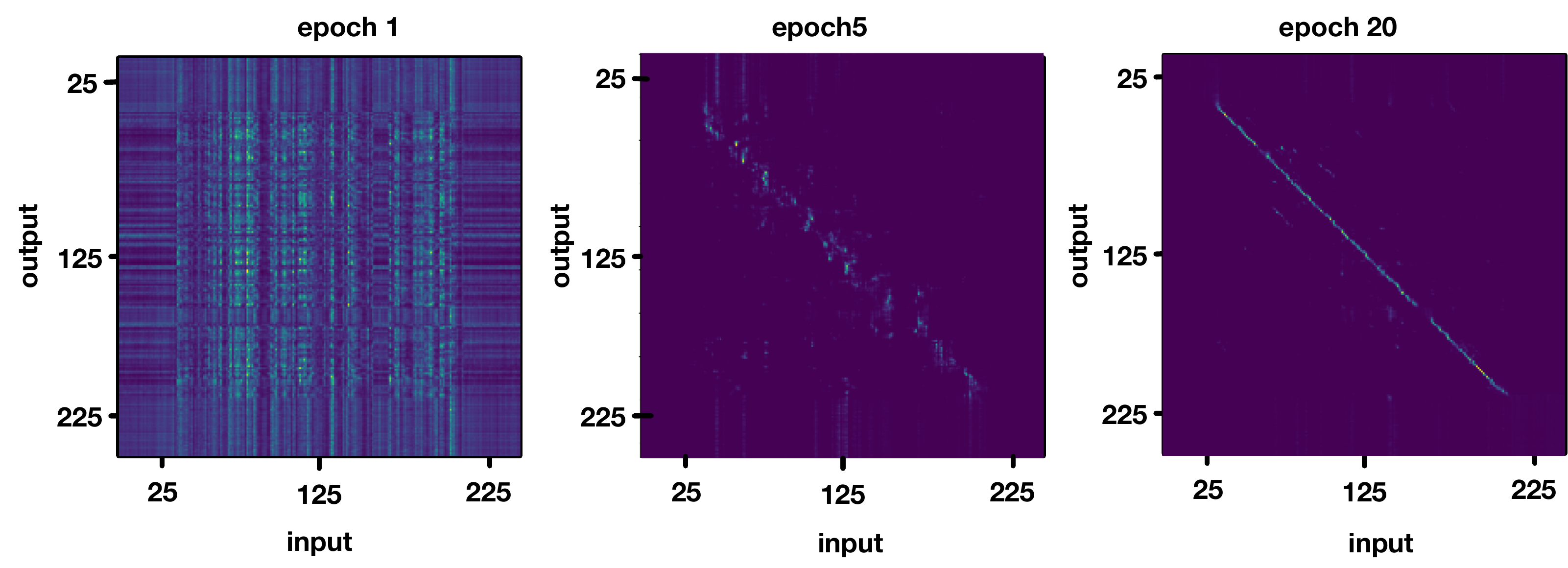}
  \caption{\rm self-attention matrix image of one head in MPE from example“4kac031f”. The horizontal axis represents input frames to the self-attention block, the vertical axis refers to the output frames of encoder.}
  \label{fig:pre_train_result}
\end{figure}

\begin{table*}[hbtp]
\caption{ \rm Results on WSJ corpus }
\centering
\begin{tabular}{cccccccc}
\toprule[1pt]
\multicolumn{1}{c}{\multirow{2}{*}{representation}} & \multicolumn{1}{c}{\multirow{2}{*}{unlabeled}} & \multicolumn{1}{c}{\multirow{2}{*}{labeled}} & \multirow{2}{*}{\begin{tabular}[c]{@{}c@{}}fine-tuning \\ steps\end{tabular}} & \multirow{2}{*}{dev93} & \multirow{2}{*}{eval92} & \multicolumn{2}{c}{baseline(supervised)} \\
\multicolumn{1}{c}{}                                & \multicolumn{1}{c}{}                           & \multicolumn{1}{c}{}                         &                                                                               &                        &                         & dev93              & eval92             \\ \hline
MPE                                                  & WSJ(81h)                                        & one-third(25h)                                & 5500                                                                          & 10.43                  & 9.31                   & 15.05              & 12.54              \\
MPE                                                  & WSJ(81h)                                       & half(40h)                                     & 3300                                                                          & 9.77                   & 7.04                    & 12.58              & 10.07              \\
MPE                                                  & WSJ(81h)                                        & WSJ(81h)                                      & 15000                                                                         & 6.79          & 4.26           & 7.93               & 5.48               \\
MPE                                                  & LibriSpeech(960h)                               & WSJ(81h)                                      & 12000                                                                         & \textbf{5.82}                     &  \textbf{3.48}                      & 7.93               & 5.48               \\ \hline
wav2vec{\cite{wave2vec}}                                        & LibriSpeech(960h)                               & WSJ(81h)                                      & -                                                                             & 6.84                   & 3.97                    & -                  & -                  \\
DeCoAR{\cite{Shaoshi2019}}                                         & LibriSpeech(960h)                               & WSJ(81h)                                      & -                                                                             & 6.30                   & 3.17                    & -                  & -                  \\
DeCoAR{\cite{Shaoshi2019}}                                         & WSJ(81h)                                        & WSJ(81h)                                      & -                                                                             & 8.34                   & 4.64                    & -                   & -                  \\
\bottomrule[1pt]
\end{tabular}
    \label{tablewsj}
\end{table*}  

\begin{table*}[htbp]
\caption{ \rm Results on LibriSpeech  corpus}
\centering
\begin{tabular}{ccccccc}
\toprule[1pt]
\multirow{2}{*}{unlabeled data} & \multirow{2}{*}{labeled data} & \multirow{2}{*}{fine-tuning steps} & \multirow{2}{*}{dev clean} & \multirow{2}{*}{test clean} & \multicolumn{2}{c}{baseline(supervised)} \\
                                &                               &                                    &                            &                             & dev clean                   & test clean \\ \hline
LibriSpeech(960h)               & train-clean-100               & 7500                               & 8.12                       & 9.68                        & \multicolumn{1}{c}{11.63}  & 12.17      \\
LibriSpeech(960h)               & train-clean-360               & 13000                              & 6.44                       & 7.83                        & \multicolumn{1}{c}{8.35}   & 9.70       \\
-                               & LibriSpeech-960               & -                                  & -                          & -                           & \multicolumn{1}{c}{3.24}   & 3.77      \\
\bottomrule[1pt]
\end{tabular}
\label{table2}
\end{table*}
In pre-training stage, in order to measure the reconstruction discrepancy, we have tried L1 loss, optimized by Adam optimizer. Figure \ref{fig:pre_train_result} shows the tendency of alignment between original frames and reconstruction frames. From left to right respectively represents the matrix image in epoch1, epoch5 and epoch20. In first epoch, the self-attention matrix image is random but gradually become orthogonal after several training epochs.  
\subsection {Supervised fine-tuning results}
In order to evaluate the two fine-tuning methods that proposed in section 2.3, a simple experiment on WSJ subset with the two methods has been implemented. Table \ref{table3} revealed frozen fine-tuning method performes better than simply initialize the encoder.
\begin{table}[h]
\caption{\rm comparison of two fine-tuning methods}
\centering
\begin{tabular}{c|c|c|c}
\toprule[1pt]
\begin{tabular}[c]{@{}c@{}}Fine-tuning \\ methods\end{tabular} & unlabeled & labeled  & dev93 \\ \hline
Directly fine-tuning                                             & WSJ(81h)  & WSJ(25h) & 14.77 \\
Frozen fine-tuning                                                 & WSJ(81h)  & WSJ(25h) & 10.43 \\
\bottomrule[1pt]
\end{tabular}
\label{table3}
\end{table}
The given results in all these tables are an average of WER in two runs. Specifically, in the decoding stage, we applied beam search (beam width=10) and CTC decoding method.
\subsubsection{Results  on WSJ}
The results of WSJ are depicted in Table \ref{tablewsj}. In WSJ corpus, for comparison, we select one-third, a half and entire data from it respectively. Firstly, we trained the three subsets on JCT structure in fully supervised manner without pre-training process. Afterwards, we trained MPE with the whole $\textit{si284}$ which contains 81h audio data without its transcription, then the three subsets are used for supervised training stage. The results shows:compared with fully supervised training, two stage training achieves 22\% wer reduction on dev93 and 30\% wer reduction on eval92. Besides, in order to test the robustness of masked pre-training method, we applied MPE which was trained by LibriSpeech-960h domain to WSJ 81h domain for parameters' fine-tuning. We can see from the table that increasing unlabeled data for MPE naturally attains better results, it achieves 22\% WER reduction in WSJ (81h).

In the bottom of Table \ref{tablewsj}, we provide the comparison of MPE and other related published representations: wav2vec, DeCoAR. Wave2vec constructs a model of five-layer convolutional neural networks. DeCoAR constructs a structure of LSTM netural networks. Compared to these two structure, we achieved 15\% wer reduction than wav2vec and 7\% wer reduction than DeCoAR on dev93. While the result on eval92 behaves not so desirable, we consider that the data set is approaching saturation and we'll propose some new idea to address this issue in future work.
\subsubsection{Results  on LibriSpeech}
Table \ref{table2} demonstrates the results of Librispeech subsets. MPE has been trained on 960 hours Librispeech unlabeled audio data, while \textit{train-clean-100} and \textit{train-clean-360} were chosen to be labeled dataset in fine-tuning stage. The supervised baseline are also given in the table. Compared with the baseline, two-stage training obtained 34\% and 25\% wer reduction on \textit{dev clean} and \textit{test clean}, respectively.
\section{Conclusion}
According to all the above experiments, two-stage training has significantly remarkable performance better than fully-supervised training. It suggests that with massive unlabeled data and limited labeled data we can achieve the same performance with the system which has been trained by a large amount of supervised data. Meanwhile, relying on the powerful modeling ability of Transformer, the masked pre-trained representation can be widely used to other down stream speech tasks.
\bibliographystyle{IEEEtran}


\end{document}